\documentclass{ws-ijmpa}

\usepackage{amsmath,amssymb}

\def\Journal#1#2#3#4{{#1} {\bf #2}, #3 (#4)}

\def\PRD{{\it Phys. Rev.}}

\newcommand{\ba}{\begin{array}}
\newcommand{\ea}{\end{array}}
\newcommand{\bea}{\begin{eqnarray}}
\newcommand{\eea}{\end{eqnarray}}
\newcommand{\be}{\begin{equation}}
\newcommand{\ee}{\end{equation}}

\newcommand{\nn}{\nonumber}
\newcommand{\lb}{\label}
\newcommand{\re}[1]{(\ref{#1})}

\newcommand{\bra}{\langle}
\newcommand{\cket}{\rangle}
\newcommand{\ve}[1]{\vec{#1}}
\begin{document}

\markboth{V.N. Pervushin et al}
{The Kinetic Description of Vacuum Particle Creation...}

%
\catchline{}{}{}{}{}
%

\title{THE KINETIC DESCRIPTION OF VACUUM PARTICLE CREATION
 IN THE OSCILLATOR REPRESENTATION}

\author{V.N. PERVUSHIN and V.V. SKOKOV}

\address{ Bogoliubov Laboratory of Theoretical Physics, Joint Institute
     for Nuclear Research,\\ 141980, Dubna, Russia}


\author{ A.V. REICHEL,  S.A. SMOLYANSKY, and  A.V. PROZORKEVICH}

\address{Physics Department, Saratov State University, 410071,
Saratov, Russia\\E-mail: smol@sgu.ru}

\maketitle


\begin{abstract}
The oscillator representation is used for the non-perturbative
description of vacuum particle creation in a strong time-dependent
electric field in the framework of scalar QED. It is shown that
the method can be more effective for the derivation of the quantum
kinetic equation (KE) in comparison with the Bogoliubov method of
time-dependent canonical transformations. This KE is used for the
investigation of  vacuum creation in periodical linear and
circular polarized electric fields and also in the case of the
presence of a constant magnetic field, including the back reaction
problem. In particular, these examples are applied for a model
illustration of some features of vacuum creation of
electron-positron plasma within the planned experiments on the
X-ray free electron lasers.

\keywords{Oscillator representation; kinetic equation;
vacuum particle creation; strong fields.}
\noindent \textit{PACS}: 11.10.Ef, 25.75.Dw
\end{abstract}

\section{Introduction}

The kinetic description of evolution of particle-antiparticle
plasma systems, created from vacuum under action of strong fields,
has formed now as an active developing branch of modern
relativistic kinetic theory. Initially, the corresponding
Schwinger term in the kinetic equation (KE) was introduced on the
semi-phenomenological basis \cite{Gur}. The first attempts to
obtain it on the dynamical level were fulfilled in
works~\cite{Rau}. The fully dynamical ground KE of such kind was
obtained in works \cite{Smol,Klu} in the framework of QED. Later,
large efforts were directed to the application of these KEs for
the description of pre-equilibrium evolution of quark-gluon
plasma, arising after collision of two ultrarelativistic heavy
ions \cite{Sch,Sch2}, as well as of the forming of
electron-positron plasma from vacuum in the planned experiments on
the X-ray free electron lasers (XFEL)
 at DESY \cite{And} and SLAC \cite{Slac}.

However, the existing procedures (Bogoliubov transformation method
\cite{Grib} in combination with some specific methods of the
kinetic theory \cite{Smol} and the formalism of the single-time
Wigner distribution functions \cite{Bia,Hoe,qke}) of the dynamical
derivation of the KE for the description of the vacuum tunneling
processes are rather complicated and limited (especially for
particles with many degrees of freedom). Originally, the kinetic
theory of such kind was constructed on the dynamical basis only
for the case of charged scalar and Dirac particles (separately
\cite{Sch} and together \cite{Sch2}) in a strong linear polarized
space homogeneous time-dependent classical electric field
(application of these QED equations to the flux tube model in the
theory of ultrarelativistic heavy ion collision may be based on
the Abelian projection method \cite{Hoo}). On the other hand,
large difficulties are met even by the attempt of the transition
from the electric field of linear polarization  to an
arbitrary one in the framework of the both 
above mentioned methods.

Having in mind the transition to the description of more realistic
systems, we are interested in looking for some alternative, more
effective formalism of the derivation of the basic KE of the
vacuum creation theory. For this aim, we try here the oscillator
(or holomorphic) representation (OR) of quantum field theory
\cite{Ef,DEGN} as a simple example of the scalar QED with a strong
electric field (Sect. 2). This formalism is found very effective
in some problem series of the quantum field theory (e.g.,
\cite{Per}). The approach based on
the Green's functions method \cite{Fradkin,Gavrilov} should be mentioned.
This method in combination with the
Kadanoff-Baym covariant formalism \cite{Maino} can lead to some alternative
variant of the kinetic description of the vacuum particle creation.

In comparison with Bogoliubov method of time-dependent transformations,
the OR leads at once to the diagonal form of the Hamiltonian
(the quasi-particle representation \cite{Grib}) and to the
specific operator equations of motion, which are the basic
elements for the derivation of the KE \cite{Smol}. The KE itself
is obtained in Sec. 3 for the space homogeneous time-dependent
quasi-classical electric field of an arbitrary polarization. As a
result of numerical solution of the KE, we show, first of all,
that the circular polarized harmonic time-dependent electric field
is more effective relatively to vacuum particle creation in
comparison with the case of the linear polarized radiation.

A more complicated case is considered in Sec. 4, where vacuum
particle creation is investigated in the presence of the
time-dependent electric field of linear polarization and the
collinear constant magnetic field. We study also the back reaction
problem, based on the union of the KE and the regularized Maxwell
equation. Since the corresponding generalization of the spinor QED
is nontrivial, the examples considered in Sec. 3 and 4 can be used
to make some qualitative predictions on the results of the planned
vacuum particle creation on the XFEL experiments. Finally, Sec. 5
sums up the results of the work.

We use the metric $ g^{\mu\nu} = diag(1,-1,-1,-1) $ and the
natural units $ \hbar = c = 1 $.

\section{The oscillator representation}

We will show as the first step that the holomorphic \cite{fs} or
the OR \cite{Ef,DEGN} leads at once to
the quasi-particle representation in the scalar QED with an
external field. So we consider here the case of a complex scalar
field in some classical space homogeneous time-dependent
electric field with 4-potential (in the Hamilton gauge)
\be\label{eq1}
    A^{\mu}(t) = \left (0,A^{1}(t),A^{2}(t),A^{3}(t) \right ),
\ee
and the corresponding field strength $\vec E=-\dot{\vec A}$, where
the overdot denotes the differentiation with respect to time.
This field can be considered either as external field or as the
result of mean field approximation based on substitution of the
quantized electric field $\tilde A^k(t)$ with its mean value,
$\langle\,\tilde A^k(t)\,\rangle =A^k(t)$, where symbol
$\langle\,\ldots\,\rangle $ denotes some averaging
operation. In the kinetic theory, taking account of fluctuations
leads to collision integrals. Thus, mean field approximation means
neglect of dissipative effects.

The Lagrange density
\be\label{eq2}
    \mathcal{L}(x) = D_{\mu}^{*} \varphi^{*} D^{\mu}
     \varphi - m^{2}\varphi^{*}    \varphi
\ee
($ D_{\mu} = \partial_{\mu} + i e A_{\mu} $, $e$ is particle
charge with its sign) leads to the equation of motion
\be\label{eq2a}
    (D_{\mu}D^{\mu} + m^2) \varphi = 0.
\ee
The space homogeneity of the system allows
to look for the
solution of the Eq. (\ref{eq2a}) in the following form
\be\label{eq2b}
    \varphi(x) = \int [dp] {\rm e}^{i \ve{p}\, \ve{x}}
\varphi(\ve{p},t),
\ee
where $ [dp] = (2\pi)^{-3/2}d^3p $. Then the equation of motion of
oscillator type follows from the Eqs. (\ref{eq2a}) and
(\ref{eq2b})
\be\label{eq2c}
    \ddot{\varphi}^{(\pm)}(\ve{p}, t) + \omega^2(\ve{p}, t)
\varphi^{(\pm)}(\ve{p},
    t) = 0
\ee
with the time-dependent frequency
\be\label{eq2d}
    \omega(\ve{p}, t) = \sqrt{m^2 + \ve{P}^2},\qquad
\ve{P}=\ve{p}-e\ve{A}.
\ee
Symbols $ (\pm) $ correspond to positive and negative frequency
solutions of the Eq. (\ref{eq2c}). We suppose a finite limit
$ \lim\limits_{t \rightarrow -\infty} A^k(t) = A^k_- $
in the infinite past and the solutions $ \varphi^{(\pm)}(\ve{p},
t) $ become asymptotically free:
\be\label{eq2e}
    \varphi^{(\pm)}(\ve{p}, t) \mathop{\rightarrow}_{t \rightarrow
\,-\infty}
    {\rm e}^{\pm i \omega_- t},\quad\mbox{where} \quad
\omega_- = \lim\limits_{t \rightarrow \,-\infty} \omega(t). \ee
The Eq. (\ref{eq2c}) is the starting point of the definition of
time-dependent frequency. Now one can introduce the decomposition
of the field functions
\begin{align}\lb{eq4}
 \varphi(x) & = \int [dp]
 \frac{{\rm e}^{i \ve{p} \, \ve{x}}} {\sqrt{2 \omega(\ve{p}, t)}}
 \left \{
 a^{(-)}(\ve{p},t) + b^{(+)}(-\ve{p},t)
 \right \}, \nn \\
 \varphi^{*}(x) & = \int\limits^{\mathstrut} [dp]
 \frac{{\rm e}^{-i \ve{p} \, \ve{x}}}    {\sqrt{2 \omega(\ve{p},t)}}
 \left \{
 a^{(+)}(\ve{p},t) +  b^{(-)}(-\ve{p},t)
 \right \}
\end{align}
and generalized momenta
\begin{align}\lb{eq5}
 \pi^{*}(x) &= i \int [dp]
 \sqrt{\frac{1}{2} \omega(\ve{p},t)} {\rm e}^{-i \ve{p} \, \ve{x}}
 \left \{
 a^{(+)}(\ve{p},t) - b^{(-)}(-\ve{p},t)
 \right \}, \nn \\
 \pi(x) &= -i \int\limits^{\mathstrut} [dp]
 \sqrt{\frac{1}{2} \omega(\ve{p},t)} {\rm e}^{i \ve{p} \, \ve{x}}
 \left \{
 a^{(-)}(\ve{p},t) - b^{(+)}(-\ve{p},t)
 \right \}.
\end{align}
The Eqs. (\ref{eq4}) and (\ref{eq5}) can be obtained from the
corresponding decompositions of the free field with the help of
formal substitution $ \omega(\ve{p}) = \sqrt{m^2 + \ve{p}\,^2}
\rightarrow \omega(\ve{p}, t) $, that is admissible for space
homogeneous case.

Thus, the decompositions (\ref{eq4}) and (\ref{eq5}) are
postulated essentially.
 It is based on the possibility of the
introduction of the canonical quantization by the complete analogy
with
 the case of the free field. Indeed, usual commutation
relations for the time-dependent creation and annihilation
operators
\be\label{comm}
 \left [ a^{(-)}( \ve{p},t), a^{(+)}( \ve{q},t) \right ] =
 \left [ b^{(-)}( \ve{p},t), b^{(+)}( \ve{q},t) \right ] =
 \delta( \ve{p} - \ve{q} )
\ee
etc. follow at once
 from the canonical
commutation relations
for the operators $ \varphi(x) $ and $ \pi(x) $ as a direct
consequence of the decompositions (\ref{eq4})
 and (\ref{eq5}).

Another feature of the OR is connected with
 the diagonal form of
the Hamiltonian in the time-dependent occupation number
representation: the substitution of the decompositions (\ref{eq4})
and (\ref{eq5}) to the full energy of the system
\be\label{eq9a}
    H_0(t) = \int d^3x \left \{
    \pi^{*}\pi + \nabla \varphi^{*} \nabla \varphi +
m^{2}\varphi^{*}\varphi
    \right \}
\ee
leads to the following Hamiltonian density in the momentum
representation
\be\label{eq6}
 H( \ve{p},t) = \omega( \ve{p},t) \left [
 a^{(+)}( \ve{p},t) a^{(-)}( \ve{p},t) +
 b^{(-)}( - \ve{p},t) b^{(+)}( - \ve{p},t) \right ]
\ee
with $ \omega(\ve{p}, t) $, defined by the Eq. (\ref{eq2d}). Thus,
the OR leads
 to the quasiparticle representation with the diagonal
Hamiltonian density (\ref{eq6}). This way is found more effective
in comparison with the "standard" one based on the procedure of
Hamiltonian diagonalization by the time-dependent Bogoliubov
transformation (e.g., \cite{Grib,niki}).

The OR can be a non-perturbative basis for the solution of one
particle problem
in the strong quasi-classical electric
field (\ref{eq1}). So let us obtain the equation of motion for the
creation and annihilation operators. For this aim, let us also
write the action
\be\label{eq8a}
 S = \int d^{4}x \left \{ \pi^{*}(x) \dot{\varphi}(x) +
 \pi(x) \dot{ \varphi^{*} }(x) -H_0(x) \right \}
\ee
in this representation
\bea\label{action}
S = \int dt\, d^{3}p \biggl\{ \frac{i}{2}\biggl(
 a^{(+)}( \ve{p},t) \dot{a}^{(-)} ( \ve{p},t) -
 \dot{a}^{(+)} ( \ve{p},t) a^{(-)}( \ve{p},t)  \nn \\ \label{eq8}
 + \dot{b}^{(-)} ( \ve{p},t) b^{(+)}( \ve{p},t) -
 b^{(-)}( \ve{p},t) \dot{b}^{(+)} ( \ve{p},t)\nn\\
 + \Delta( \ve{p},t) \bigl[
 b^{(-)}( \ve{p},t) a^{(-)}( - \ve{p},t) -
 a^{(+)}( \ve{p},t) b^{(+)}( - \ve{p},t) \bigr]\biggr) -
 H_0( \ve{p},t)
 \biggr\}, \eea
where
\be\label{ampl}
\Delta( \ve{p},t)= \frac{ \dot{ \omega }( \ve{p},t) }{ \omega(
\ve{p},t)}.
\ee
The "anomalous" terms are presented
in the Eq.(\ref{action}) in the square brackets.
These terms describe creation and annihilation of the particle-antiparticle pairs
with the vacuum transition amplitude \re{ampl} under action of the electric  field.
Then the operator equations of
motion  \cite{Grib} follow from that after variations by respect
to the amplitudes $a^{(\pm)}$, $b^{(\pm)}$ and subsequent
transition to the occupation number representation with the
commutation relations \re{comm}:
\bea\label{eq10}
 \dot{a}^{(\pm)}( \ve{p},t) &=& \frac{1}{2}\,
 \Delta( \ve{p},t)
 b^{(\mp)}( - \ve{p},t) + i \left [ H_0(t), a^{(\pm)}( \ve{p},t)
\right ], \nn
 \\
 \dot{b}^{(\pm)}( \ve{p},t) &=&\frac{1}{2}\,
\Delta( \ve{p},t)
 a^{(\mp)}( - \ve{p},t) + i \left [ H_0(t), b^{(\pm)}(
 \ve{p},t) \right ].
\eea
It is assumed that the electric field is switched off in the
infinite past
\be\label{eq11}
 \lim\limits_{t \rightarrow -\infty } \ve{E}(t) =
-\lim\limits_{t \rightarrow -\infty }\dot{\ve{A}}(t)=0, \ee
but the infinitely distant past and future asymptotics of the
vector field can be distinct
\cite{Grib,niki}, i.e.
\be\label{eq12}
 \lim\limits_{t \rightarrow - \infty } \ve{A}(t) = \ve{A}_{-} \neq
 \lim\limits_{t \rightarrow \infty } \ve{A}(t) = \ve{A}_{+}.
\ee
Let us also notice that the equations of motion (\ref{eq10})
contain some elements of generalization in comparison with the
known case \cite{Grib,pop}, where the field polarization remains
fixed. This circumstance will be used later on in Sec. 3 in order to
derive the KE for the description of vacuum particle creation in
the time-dependent electric field of arbitrary polarization. We
mark out one feature of the Heisenberg equations of motion
(\ref{eq10}): each of these equations describes some mixture of
positive and negative energy states ( "non-diagonal" form of
equations of motion). That is reflected also in the action
(\ref{eq8}), where the "anomalous" contributions $ \sim a^{(\pm)}
b^{(\pm)} $ are present. In this sense the OR leads to mixed
states.

The Eqs. \re{eq10} show, that the operator (\ref{eq6}) is not
 a generator of
infinitesimal time transformations in the considered Fock space,
i.e. it is not the "proper" Hamiltonian of the system in this
representation. However, the action (\ref{eq8}) is a homogeneous
square form relative to the operators $ a^{(\pm)} $ and $
b^{(\pm)} $ and its anomalous terms can be eliminated with the
help of some special time-dependent Bogoliubov transformation
\bea\label{eq22}
 a^{(-)}( \ve{p},t) &=&
 \alpha_{ \ve{p}}^{*}(t)\,
 c^{(-)}( \ve{p},t)\ -
 \beta_{ \ve{p}}(t)
 d^{(+)}( - \ve{p},t), \nn \\
 b^{(-)}( \ve{p},t) &=&
 \alpha_{ - \ve{p}}^{*}(t)
 d^{(-)}( \ve{p},t) -
 \beta_{ - \ve{p}}(t)
 c^{(+)}( - \ve{p},t)
 \eea
with the additional relation $ | \alpha_{\ve{p}}(t) |^{2} - |
\beta_{\ve{p}}(t) |^{2} = 1 $ (the condition of the transformation
reversibility and the form-invariance of the canonical commutation
relations (\ref{comm})). The substitution of the Eqs. (\ref{eq22})
into the Eqs. (\ref{eq10})
generates to the equations of motion for
Bogoliubov coefficients \cite{Grib,pop}:
\begin{equation}\label{eq14}
  \dot{\alpha}_{\ve{p}}(t) =\frac{1}{2}\,
\Delta( \ve{p},t)
  \beta^{*}_{\ve{p}}(t) {\rm e}^{2i \vartheta(\ve{p}; t, t_0)},
  \,\,\,\,\,
  \dot{\beta}_{\ve{p}}(t) =\frac{1}{2}\,
\Delta( \ve{p},t)
  \alpha^{*}_{\ve{p}}(t) {\rm e}^{2i\vartheta(\ve{p}; t, t_0)},
\end{equation}
where ($t_0$ is a field switching on time)
\begin{equation}\label{phase}
  \vartheta(\ve{p}; t, t_0) = \int\limits_{t_0}^t dt' \omega(\ve{p},t').
\end{equation}
Then the action (\ref{eq8}) is transformed to the standard form
\bea\label{eq25}
 S = \int dt \, d^{3}p \biggl\{ \frac{i}{2} \Bigl[ c^{(+)}(\ve{p},t)
\dot{c}^{(-)}( \ve{p},t) -
 \dot{c}^{(+)}(\ve{p},t) c^{(-)}( \ve{p},t) \nn \\+\,
 \dot{d}^{(-)}( \ve{p},t) d^{(+)}( \ve{p},t) -
 d^{(-)}( \ve{p},t) \dot{d}^{(+)}( \ve{p},t) \Bigr] -
H'(\ve{p},t) \biggr\},
\eea
where the "true"
 Hamiltonian $ H'(\ve{p},t) $ has now a
non-diagonal form in the momentum representation
(the explicit form of $H'(\ve{p},t) $ can be found in \cite{Grib}).
The equations of motion (\ref{eq10}) turns into
the usual Heisenberg equations, i.e.
\be\label{eqHeis}
 \dot{c}^{(\pm)}( \ve{p},t) =
 i \left [ H'(t), c^{(\pm)}( \ve{p},t) \right ],
\ee
where the Hamiltonian $ H'(\ve{p},t) $ is the original point of the theory
based on decomposition of the type \re{eq4} with the free particle frequency
$\omega_0(\ve{p})=\sqrt{m^2+\ve{p}^2}$.
Thus, the OR can be considered as the "turned inside out"
Bogoliubov method of the Hamiltonian diagonalization,

On the other hand, the Bogoliubov transformation of the type
(\ref{eq22}) can be used to restore the diagonal form of the
equations of motion (\ref{eq10}) without mixing the terms (the
first addends)  \cite{Per}. Then we obtain the following equations of
motion in momentum representation:
\be
i \dot{\mathcal{A}}^{(\pm)}(\ve{p},t) = \pm \Omega(\ve{p},t)
\mathcal{A}^{(\pm)}(\ve{p},t),
\ee
where $\mathcal{A}^{(\pm)}(\ve{p},t)$ are creation and annihilation
operators in the new representation with some new frequency
$\Omega (\ve{p},t)$. These operators allows us to gain
the first integral  of motion
\be\label{1int}
\bra in|\mathcal{A}^{(+)}(\ve{p},t)
 \mathcal{A}^{(-)}(\ve{p},t)|in\cket
 = \bra in|\mathcal{A}^{(+)}(\ve{p},0) \mathcal{A}^{(-)}(\ve{p},0)|in\cket ,
\ee
where the averaging procedure is fulfilled over the in-vacuum state.
The details connected with that type representation can be found
in the work \cite{Per}.
It is important that the correlator \re{1int}
can not be interpreted in quasiparticle terms \cite{Grib}.

\section{Particle creation in the time-dependent electric field}

As an example, we will obtain the KE describing vacuum scalar
particle creation in the electric field (\ref{eq1}) of arbitrary
polarization. Let us introduce the distribution function of
particles using the quasiparticle representation (see Eq.\re{eq6})
\cite{Grib}
\be\label{eqn14}
 f( \ve{p},t) = \bra in|
 a^{(+)}( \ve{p},t) a^{(-)}( \ve{p},t) |in \cket,
\ee
which is related with the distribution function of anti-particles as
$\overline{f}( -\ve{p},t)=f( \ve{p},t)$.

Using the method of works \cite{Smol} and the basic equations of
motion (\ref{eq10}), it is not difficult to get the KE.
Differentiating the distribution function  (\ref{eqn14}), we obtain
\be\label{eqn14a}
 \dot{f}(\ve{p},t) =\frac12 \Delta(\ve{p},t) \left[f^{(+)}(\ve{p},t) +
 f^{(-)}(\ve{p},t)\right],
\ee
where the anomalous one-particle correlators are
\bea\label{eqn14b}
 f^{(+)}(\ve{p},t) &=& \bra in|
 a^{(+)}( \ve{p},t)\, b^{(+)}( -\ve{p},t) |in \cket,\nn\\
 f^{(-)}(\ve{p},t) &=& \bra in|
 b^{(-)}( -\ve{p},t)\, a^{(-)}( \ve{p},t) |in \cket
\eea
and the vacuum transition amplitude is
\be\label{eqn14c}
\Delta(\ve{p},t) = \frac{e\ve{E}(t)\ve{P}(t)}
 {\omega^2(\ve{p},t)}.
\ee
The equations of motion for the functions (\ref{eqn14b}) can be
obtained by analogy with the Eq. (\ref{eqn14a}). We write them out
in the integral form
\be\label{eqn14d}
 f^{(\pm)}(\ve{p},t) =\frac12 \int\limits_{-\infty}^{t} dt' \Delta(\ve{p},t')
 \left[ 1 + 2f(\ve{p},t') \right] e^{\pm 2i\theta(\ve{p};t,t')},
\ee
where the asymptotic conditions have been introduced
\be\label{eqn14e}
 \lim_{t \rightarrow -\infty } f^{(\pm)}(\ve{p},t) = 0.
\ee
The consequence of the Eqs. (\ref{eqn14a}) and (\ref{eqn14d}) is
\be\label{eqn14f}
 \dot{f}(\ve{p},t) = \Delta(\ve{p},t) Ref^{(+)}(\ve{p},t).
\ee
Finally, the substitution of the Eqs. (\ref{eqn14d}) into the Eq.
(\ref{eqn14f}) leads to the resulting KE
\be\label{eq13}
\dot f(\ve{p},t)= I( \ve{p},t),
\ee
where the function $ I( \ve{p},t) $ is the vacuum source term of
particles
\be\label{eqn15}
 I( \ve{p},t) = \frac{1}{2} \Delta ( \ve{p},t)
 \int\limits_{-\infty}^{t} dt' \Delta (\ve{p},t') [ 1+ 2
 f(\ve{p},t') ] \cos[ 2\, \theta (\ve{p};t,t') ].
\ee
The KE (\ref{eq13}), (\ref{eqn15}) is the non-perturbative result
for the electric field (\ref{eq1}) in the mean field
approximation. The particular case of this KE for the linear
polarization was obtained and researched in detail in works
\cite{Smol,Sch}.

The KE (\ref{eq13}) can be transformed to a system of ordinary
differential equations, which is convenient for numerical analysis
\cite{Sch} (to compact our formulas here and below, symbols
denoting time- and momentum-dependency are dropped in obvious
cases)
\bea\label{ode}
  \dot{f} &=& \frac{1}{2} \Delta v_1, \nn \\
  \dot{v}_1 &=& \Delta (1 + 2f)- 2 \omega v_2, \nn\\
  \dot{v}_2 &=& 2 \omega v_1,
\eea
where $\Delta$ is defined by Eq. (\ref{eqn14c}) and $v_1 = \frac12\mathrm{Re}
f^{(+)}$, $v_2 = \frac12\mathrm{Im} f^{(+)}$.
The Eqs. (\ref{ode}) must be considered with the zero initial
conditions. This equation system has the first integral
\be\label{f_int}
    -(1+2f)^2+v_1^2+v_2^2=1,
\ee
according to it the phase trajectories are located on two-cavity
hyperboloid with top coordinates $f=v_1=v_2=0$ and $f=-1,
v_1=v_2=0$.

Excluding the function $f$ from Eqs.  \re{ode}, we obtain the two
dimensional dynamical system with the equations of motion
\bea\label{two}
   \dot v_1&=&\Delta\sqrt{1+v_1^2+v_2^2}-2\omega v_2,\nn\\
    \dot v_2&=&2\omega v_1^{\phantom{\mathstrut}},
 \eea
having non-Hamilton structure. This is the direct consequence
of unitary nonequivalence between the Heisenberg equations of motion
\re{eqHeis} (Hamilton dynamics) and the Heisenberg type equations
\re{eq10}, which is the basis of the KE \re{eq13} and the system
\re{two} (non-Hamilton dynamics).

Let us remark also, that the full equation set \re{ode} have no
reversal time symmetry but the physical observables are defined by
the functions $f,v_1$ only, which are not changed at time
inversion.

The system \re{ode} is integrated via the Runge-Kutta method with
the zero initial conditions
$f(\ve{p},t_0)=v_1(\ve{p},t_0)=v_2(\ve{p},t_0)=0$. The momentum dependence
of distribution function is defined
 by means of digitization of
momentum space to 2 or 3-dimensional grid, in each of its node the
system \re{ode} is solved.
 The concrete grid parameters depend on
field strength, the typical values are $\Delta p\approx 0.1m$
(grid step) and $p_{max}\approx (10-15)m$ (grid boundary), so the
total number of solved equations is of order $10^5-10^6$.

Now we use the KE (\ref{eq13})  to compare the effectiveness of
particle creation in the harmonic time-dependent electric field of
circular ($\xi =1$) polarization and of linear ($\xi =0$) one
(the analog of last case was considered  in the spinor QED in
works \cite{And})
\begin{equation}\label{field}
  A^\mu (t) = (0,0,\xi \chi(t) A\cos{\nu t}, \chi(t) A\sin{\nu t}),
\end{equation}
where $\chi(t)$ is some function, providing switching on of the
field smoothly on a half of the field period (the result does not
depend on the explicit form of the switching function when $t\gg
1/\nu$). Time dependence of particle number density
\begin{equation}\label{den}
    n(t)=\int [dp] f(\ve{p},t)
\end{equation} for the fields
(\ref{field}) with the zero initial conditions at $t_0=0$
is shown in Fig. 1
(in the natural units). We have chosen the field strength
parameter typical for the planned experiments on XFEL \cite{And},
but the considerably larger frequency $\nu=0.1m$
(this is convenient for calculation time reduction).

As it follows from Fig. 1,
circular polarized field is more effective for vacuum particle
creation in comparison with linear polarized field of the same
amplitude. The dynamics of particle creation in periodic field
changes qualitatively by the increase of the field strength within
the range $(0.1-0.5) E_{cr}$, where $E_{cr}=m^2/e$. Actually, when
$E=0.1 E_{cr}$ the time dependence of particle density is periodic
with the double frequency and approximately constant mean density
 value for the period of laser field (Fig. 1, left panel).
On the contrary, the mean density increases with time when $E=0.5
E_{cr}$ (the density "accumulation" effect \cite{And}, Fig. 1,
right panel).
This effect is more expressed  in the circle polarized field. We
prove that for our toy model  the creation rate (of charged
bosons) in the circle polarized field $\approx$ 4 times larger
than that of in the linear polarized field. We think that
this proper effect can be valid for the conditions of the planned
experiments on the XFEL.

\begin{figure}
\centering
\includegraphics[width=55mm,height=45mm]{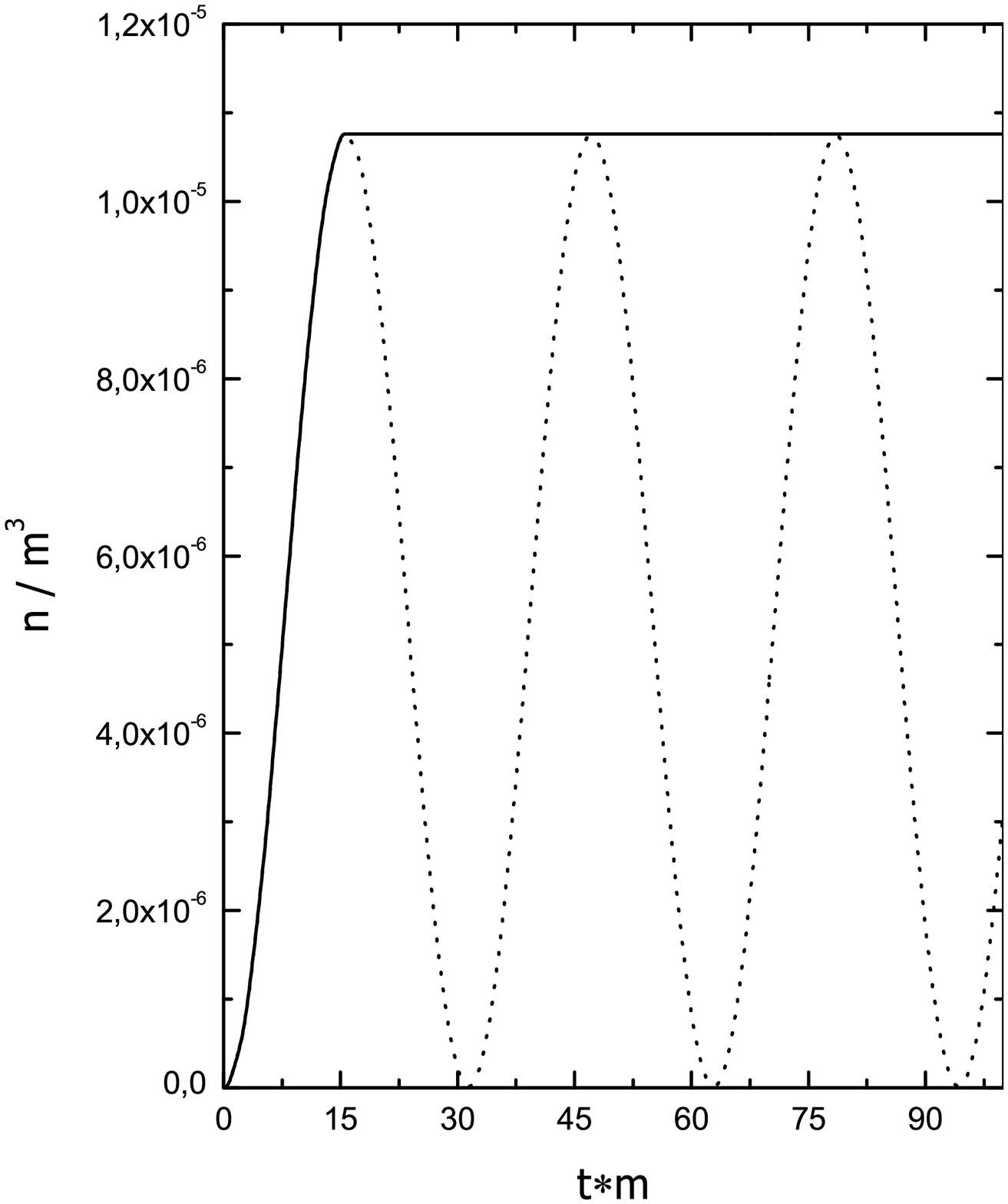}\hspace{5mm}
\includegraphics[width=55mm,height=45mm]{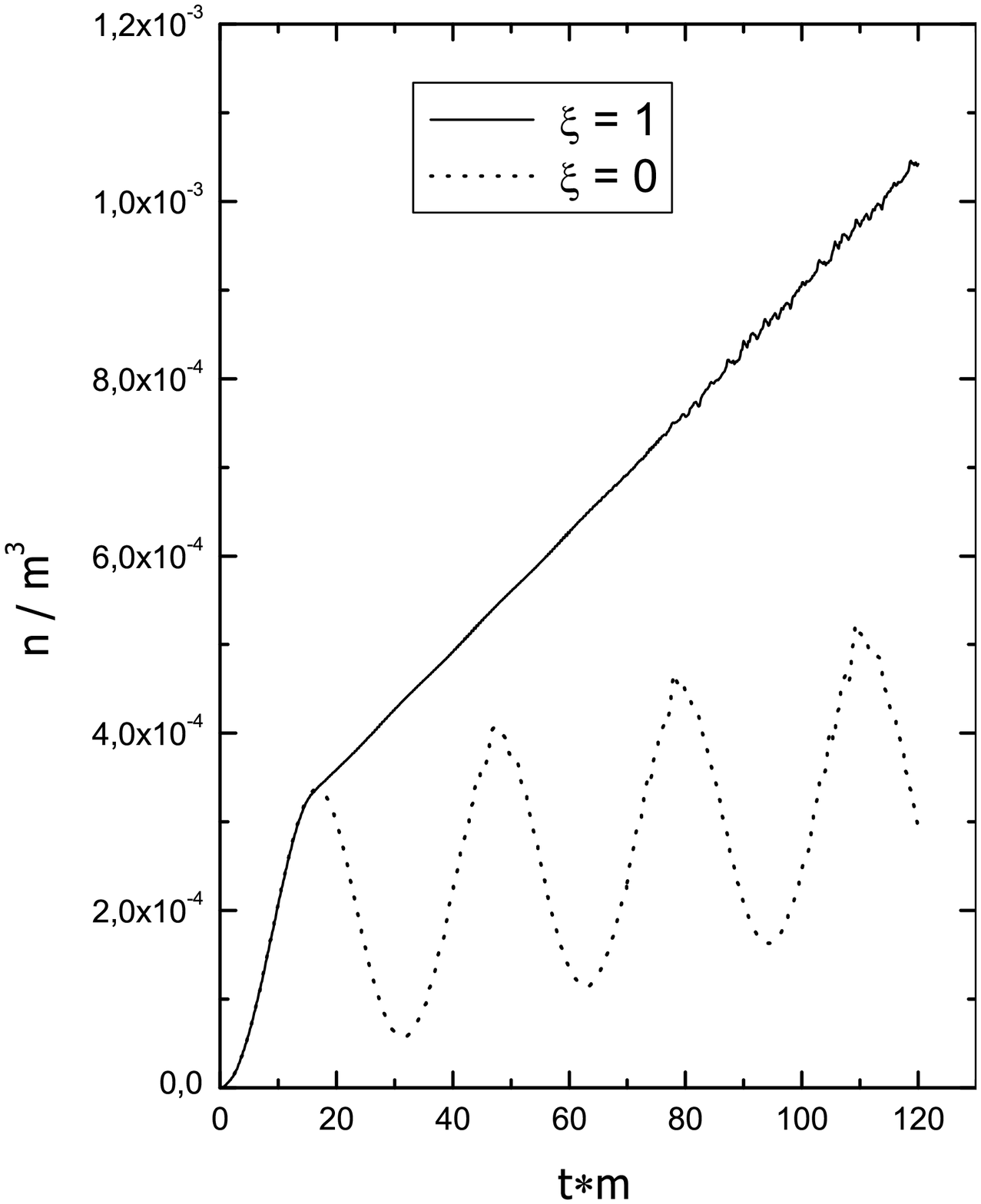}
\vspace*{8pt}
\caption{Time dependence of particle number density for the cases of
linear and circular polarized field : $E=0.1E_{cr}$ (left panel),
$E=0.5E_{cr}$ (right panel). The solid lines correspond to circular
 polarization
($\xi=1$) and dotted lines correspond to linear one ($\xi=0$).}
\end{figure}

If the field strength is of the order  of critical one, it is
necessary to take into account the back reaction
of the produced plasma on the primary field. Then the external
field $\ve{E}(t)$ should be replaced by the total field
\begin{equation}\label{eqn33}
    \ve{E}(t) = \ve{E}_{in}(t) + \ve{E}_{ex}(t),
\end{equation}
where $\ve{E}_{ex}(t)$ is known function and $\ve{E}_{in}(t)$
obeys the Maxwell equation
\begin{equation}\label{eqn34}
    \frac{d\ve{E}_{in}}{dt} = -e \int\limits [dp]
    \frac{\ve{p}}{\omega} (2f + v_1).
\end{equation}
The total current density
in the r.h.s. of this equation is the sum of conductivity and
vacuum polarization currents. Momentum dependency of the functions
$f(\ve{p},t)$ and $v_1(\ve{p},t)$ is implicit and it is necessary to
investigate their ultraviolet behavior in order to fulfill the
regularization of the integral (\ref{eqn34}). The method of the
asymptotic expansion can be most adequate one in the framework of
the considered non-perturbative formalism. The regularization
procedure proposed below is a variant of n-wave regularization
technique \cite{nreg}. Calculation details can be found in the
Appendix A.
The regularisation  in the r.h.s. of the
Eq. (\ref{eqn34}) can be achieved by substitution $v_1\to v_1-v^c_1$, where
the counter-term $v_1^c(\ve{p},t)$ is the leading term of the asymptotic series of
the function $v_1(\ve{p},t)$ over the inverse powers of momentum
$|\ve{p}|$
\begin{equation}\label{eqn35}
    v_1^c(\ve{p},t) = \frac {e\dot{\ve{E}}(t) \ve{p}}
    {4|\ve{p}|(p^2+M^2)^{3/2}},
\end{equation}
where  $M$ is an auxiliary  mass parameter, $M \gg
m$ (some modification of this procedure leads
to charge renormalization in the Eq. \re{eqn34} \cite{prepr,Kluger}).
The counter terms  become significant only in the ultraviolet
region. One can choose $M \gg
\Lambda$, where $\Lambda$ is a grid boundary of numerical
calculation of the integral (\ref{eqn34}). That allows to omit
counter-terms as a negligible quantity. The KE (\ref{eq13})
(or the Eqs. \re{ode}) and
the Maxwell equation (\ref{eqn34}) form
 the complete equation
set of the back reaction problem, which have been investigated
initially in
works \cite{Sch} for the case of soliton-like pulse of the
external linear polarized field.
It was shown that the action of the sufficiently strong
 field $E > E_{cr}$ excites
 the plasma oscillations with a period, dependent
on the residual density after termination of the external pulse.
In the considered case $E < E_{cr}$, the density of excited plasma
is small and the effects of back reaction are negligible.

\section{Magnetic field influence}

Now we shall investigate the vacuum creation problem in the case
of the collinear time-dependent electric and constant magnetic
fields
\begin{equation}\label{eqn36}
    \ve{E}(t) = (0,0,E(t)), \qquad
    \ve{H} = (0,0,H).
\end{equation}
That corresponds to the following configuration of vector
potential in the Hamilton gauge
\begin{equation}\label{eqn37}
    A^{\mu}(\ve{x},t) = (0, -Hx_1,0, A(t)).
\end{equation}
Some preliminary results  were reported on the conference
\cite{Dre}.

In general case, the introduction of the OR is based on the
possibility to define the corresponding dispersion law
$\omega(\ve{p},t)$ in the presence of the electromagnetic field.
This is possible when the spatial and time variables are
separable, in particular,  for the Klein-Gordon equation
(\ref{eq2a}) in the field (\ref{eqn36}). That leads
 to the following solutions
\begin{equation}\label{eqn38}
    \varphi^{(\pm)}_{n,p_2,p_3}(x) =
C_nT_n^{(\pm)}(t)e^{-\eta^2/2}H_n(\eta)e^{i(p_2x_2+p_3x_3)}\equiv
    T_n^{(\pm)}(t) \Phi_{n,p_2,p_3}(x),
\end{equation}
where $H_n(\eta)$ are the Chebyshev-Hermite polynomials, $\eta =
\sqrt{|e|H} (x_1+p_2/|e|H)$ and
\begin{equation}\label{eqn39}
    C_n =
\frac{1}{2\pi}\left(\frac{1}{\pi}|e|H\right)^{1/4}(2^nn!)^{-1/2}
\end{equation}
is a normalizing constant. The functions $T_n^{(\pm)}(t)$ obey the
oscillator-type  equation
\begin{equation}\label{eqn40}
    \ddot{T}_n^{(\pm)} + \omega_n^2(t) \,T_n^{(\pm)}=0
\end{equation}
with the dispersion law ($P = p - eA(t)$, $p = p_3$)
\begin{equation}\label{omgn}
    \omega_n^2(p,t) = m^2+P^2+|e|H(2n+1), \;\; n = 0,1,...
\end{equation}
The symbols $(\pm)$ over
 the functions $T_n^{(\pm)}$ correspond to
positive and negative frequency solutions of the Eq. (\ref{eqn40})
at $t\rightarrow-\infty$ like in Sect. 3. Finally, the solution of
of the Eq. (\ref{eqn40}) satisfies the relation
\begin{equation}\label{eqn42}
    iT_n^{*(\pm)}(t) \mathop{\partial_t}^\leftrightarrow
    T_n^{(\pm)}(t)=\mp1,
\end{equation}
as a consequence  of the general orthonormalization
 condition for the solutions of the Eq. (\ref{eq2a})
\begin{equation}\label{eqn43}
    i\int d^3x
    \varphi^{*(\pm)}_{n,p_2,p}(x)\mathop{\partial_t}^\leftrightarrow
    \varphi^{(\pm)}_{n',p'_2,p'}(x) =
    \mp\delta_{nn'}\delta(p_2-p'_2)\delta(p-p').
\end{equation}
The oscillator decomposition for the field functions and
generalized momenta can be constructed on the basis of the
function set (\ref{eqn38}) by analogy with the Eqs. (\ref{eq4})
and (\ref{eq5})
\bea\label{eqn44}
    \varphi(x) = \sum_n \int
\frac{dp_2dp}{\sqrt{2\omega_n(p,t)}}
    \nn \\
    \times\left\{a^{(-)}(n,p_2,p;t)\Phi_{n,p_2,p}(x)+
    b^{(+)}(n,p_2,p;t)\Phi_{n,p_2,p}^*(x) \right\}, \nn \\
    \pi^*(x) = i \sum_n \int dp_2dp
    \sqrt{\frac{1}{2}\omega_n(p,t)} \nn \\
    \times\left\{a^{(+)}(n,p_2,p;t)\Phi_{n,p_2,p}^*(x)-
    b^{(-)}(n,p_2,p;t)\Phi_{n,p_2,p}(x) \right\}.
\eea
The substitution of the Eqs. (\ref{eqn44}) into the action
(\ref{eq8}) leads
 to the equations of motion
\bea\label{eqn45}
    \dot{a}^{(\pm)}(n,p_2,p;t) = \frac{1}{2} W_n(p,t)
    b^{(\mp)}(n,-p_2,-p;t)
    + i[H_0(t), a^{(\pm)}(n,p_2,p;t)],
    \nn \\
    \dot{b}^{(\pm)}(n,p_2,p;t) = \frac{1}{2} W_n(p,t)
    a^{(\mp)}(n,-p_2,-p;t)
    + i[H_0(t), b^{(\pm)}(n,p_2,p;t)],
\eea
where the vacuum transition amplitude is equal
\begin{equation}\label{eqn46}
    W_n(p,t) = \frac{|e|E(t)P}{\omega_n^2(p,t)},
\end{equation}
and $H_0$ is the Hamiltonian in the diagonal quasiparticle
representation  with the energy (\ref{omgn}). Let us introduce now
the new distribution function of scalar particles with momentum
$(p_2, p)$ on the Landau $n$-level
\begin{equation}\label{eqn47}
    f_n(p_2,p,t) = \bra in|
    a^{(+)}(n,p_2,p;t)a^{(-)}(n,p_2,p;t)
    |in \cket.
\end{equation}
The resulting KE now has the form
\bea\label{eqn48}
    \dot{f}_n (p,t) = \frac{1}{2}W_n(p,t)
\int\limits_{-\infty}^{t}
    dt' W_n(p,t') [1+2f_n(p,t')] \cos\{ 2\int\limits_{t'}^td\tau
\omega_n(p,\tau)\}. \eea
It was taken into account here that the distribution function does
not depend on  $p_2$ momentum projection $    f_n(p_2,p,t) =
f_n(p,t)$ due to the absence of the $p_2$-dependence of the
amplitude (\ref{eqn46}). This is the well known effect of the
degeneracy of the observed quantities on the Landau level
\cite{Stat}. We can introduce  the effective mass $m^* = \sqrt{m^2
+ |e|H} \geq m$ in the Eq. (\ref{omgn}), which leads to  the
increase of the the effective energy gap width and the suppression
of vacuum particle creation. This result is in agreement with the
known result for a particular case of constant electric and
magnetic fields combination \cite{niki,pop}.

The KE (\ref{eqn48}) can be reduced to the ordinary differential
equation system as in Sect. 3
\bea\label{eqn50}
    \dot{f}_n &=& \frac{1}{2} W_n v_{1n}, \nn\\
    \dot{v}_{1n} &=& W_n (1+2f_n) - 2\omega_n v_{2n}, \\
    \dot{v}_{2n} &=& 2\omega_n v_{1n}. \nn
\eea

The solution of the back reaction problem is based on the KE
(\ref{eqn48}) and the Maxwell equation (the factor $|e|H$ is the
consequence of the above mentioned degeneracy \cite{Stat})
\begin{equation}\label{eqn51}
    \frac{dE_{in}}{dt} = -2e
    \frac{|e|H}{(2\pi)^2}\sum_{n=0}^\infty\int\limits_{-\infty}^\infty dp
    \frac{P}{\omega_n(p,t)} \left[f_n(p,t) + \frac{1}{2}
    v_{1n}(p,t)\right].
\end{equation}

We investigate the back reaction effect at the presence of the
critical magnetic field $H=H_{cr}=m^2/e=4.4\cdot 10^{13}Gs$ for
the initial electric pulse of the "rectangular" form
\begin{equation}\label{rect}
E_{ex}(t)=(E_m/2)[\tanh(t+b)-\tanh(t-b)]
\end{equation}
with the amplitude $E_m$ and the duration of $2b$. It is known
\cite{Grib}, that constant magnetic field suppress boson creation
and enhance fermion creation in the presence of the constant
electric field. The situation varies essentially
 for the case of
fast time-dependent electric field. Figs. 2 and 3  show the
evolution of the total electric field \re{eqn33} and particle
number density
\begin{equation}\label{dens}
  n(t) = \frac{|e|H}{(2\pi)^2}
  \sum_{n=0}^{\infty}\ \int\limits_{-\infty}^{\infty}f_n(p,t)\,dp
\end{equation}
for the
weak and strong electric field with $H=H_{cr}$ and $e^2=10$,
here $e$ is the effective colour charge
(charge value selection is motivated by the flux-tube model in the
theory of the QGP generation at ultra-relativistic heavy ion
collision~\cite{Sch,Sch2}). Total field almost coincides with
external pulse with $E_m=0.1E_{cr}$, i.e. the back reaction is
negligible here due to the small particle density. The internal
field becomes appreciable when $E_m=0.5E_{cr}$ and stable plasma
oscillations are excited after the termination of the external
pulse. The period of oscillations decreases with the growth of the
residual particle density as $\sqrt{n}$. The amplitude of the
oscillations becomes large-scale modulated at the further
increase of the field, Fig. 3.

\begin{figure}
\centering
\includegraphics[width=55mm,height=45mm]{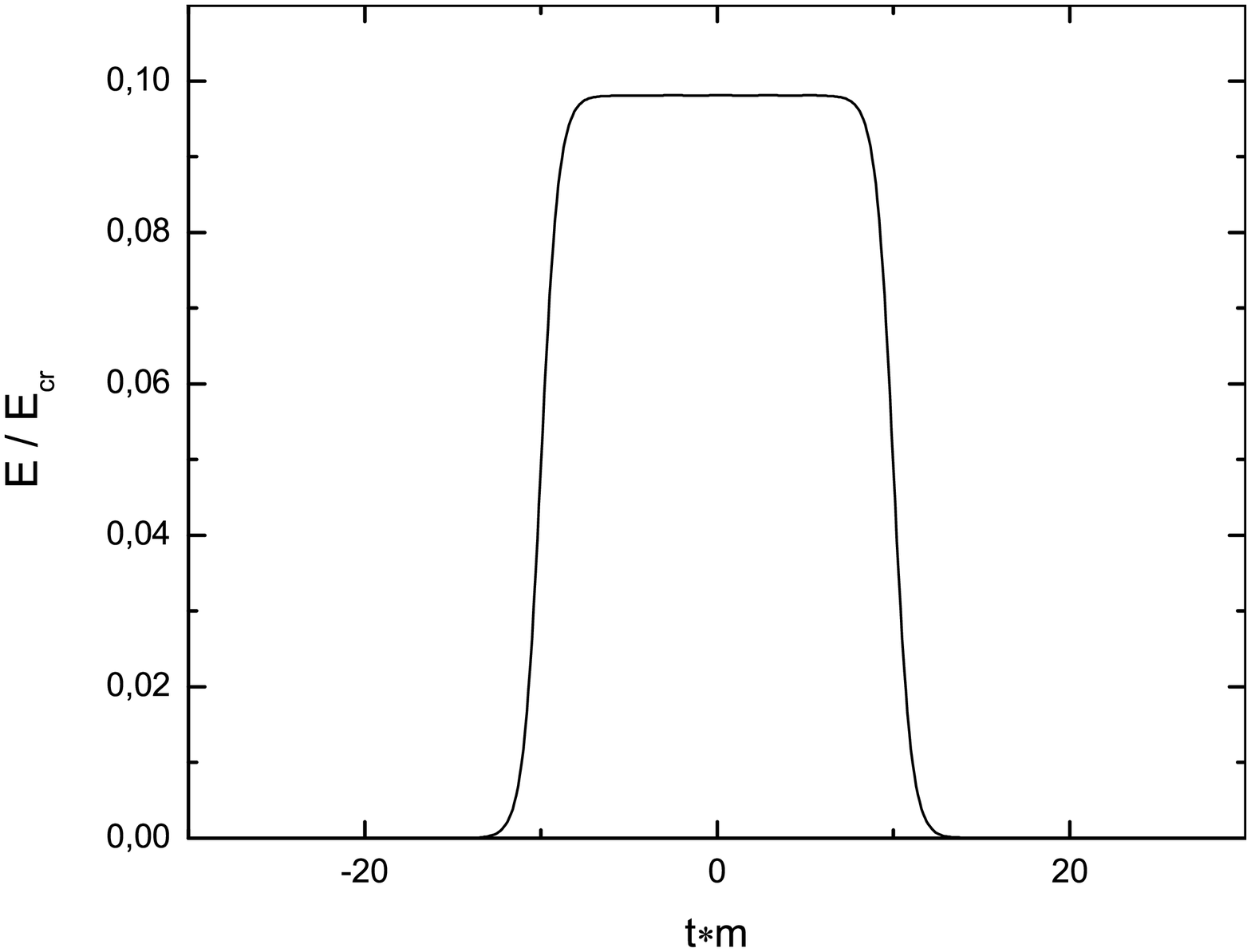}\hspace{5mm}
\includegraphics[width=55mm,height=45mm]{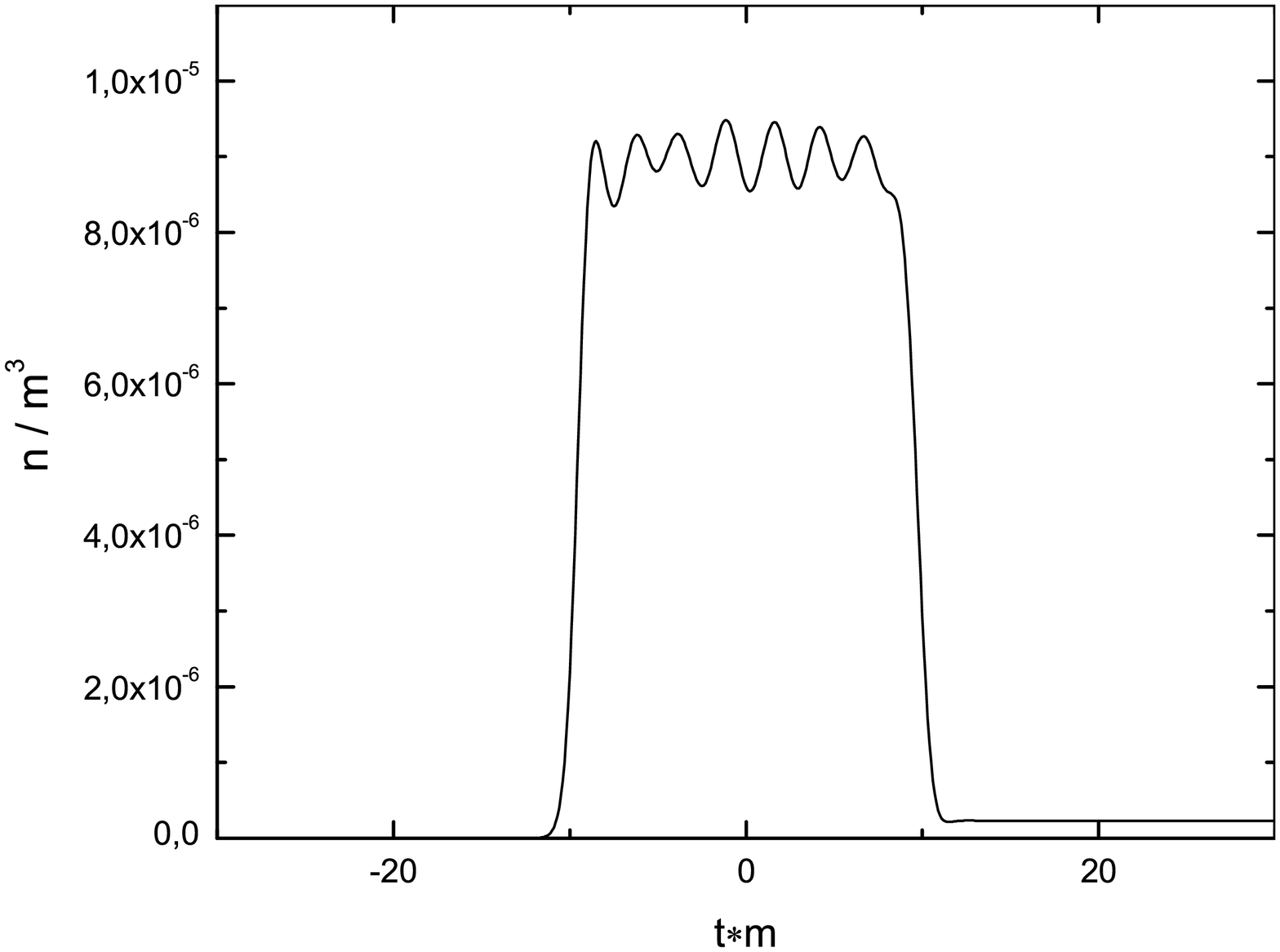}\vspace*{16pt}
\includegraphics[width=55mm,height=45mm]{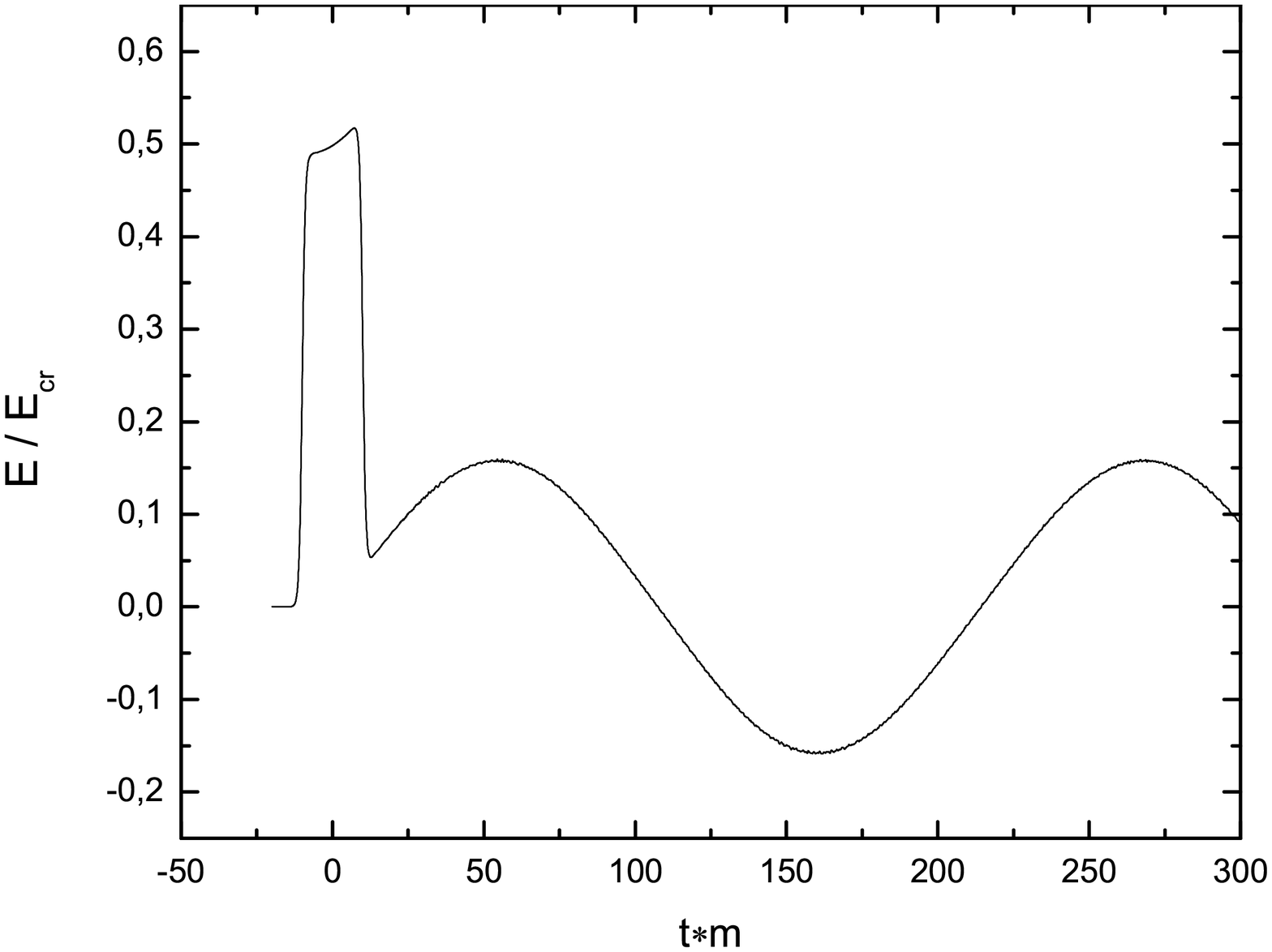}\hspace{6mm}
\includegraphics[width=55mm,height=45mm]{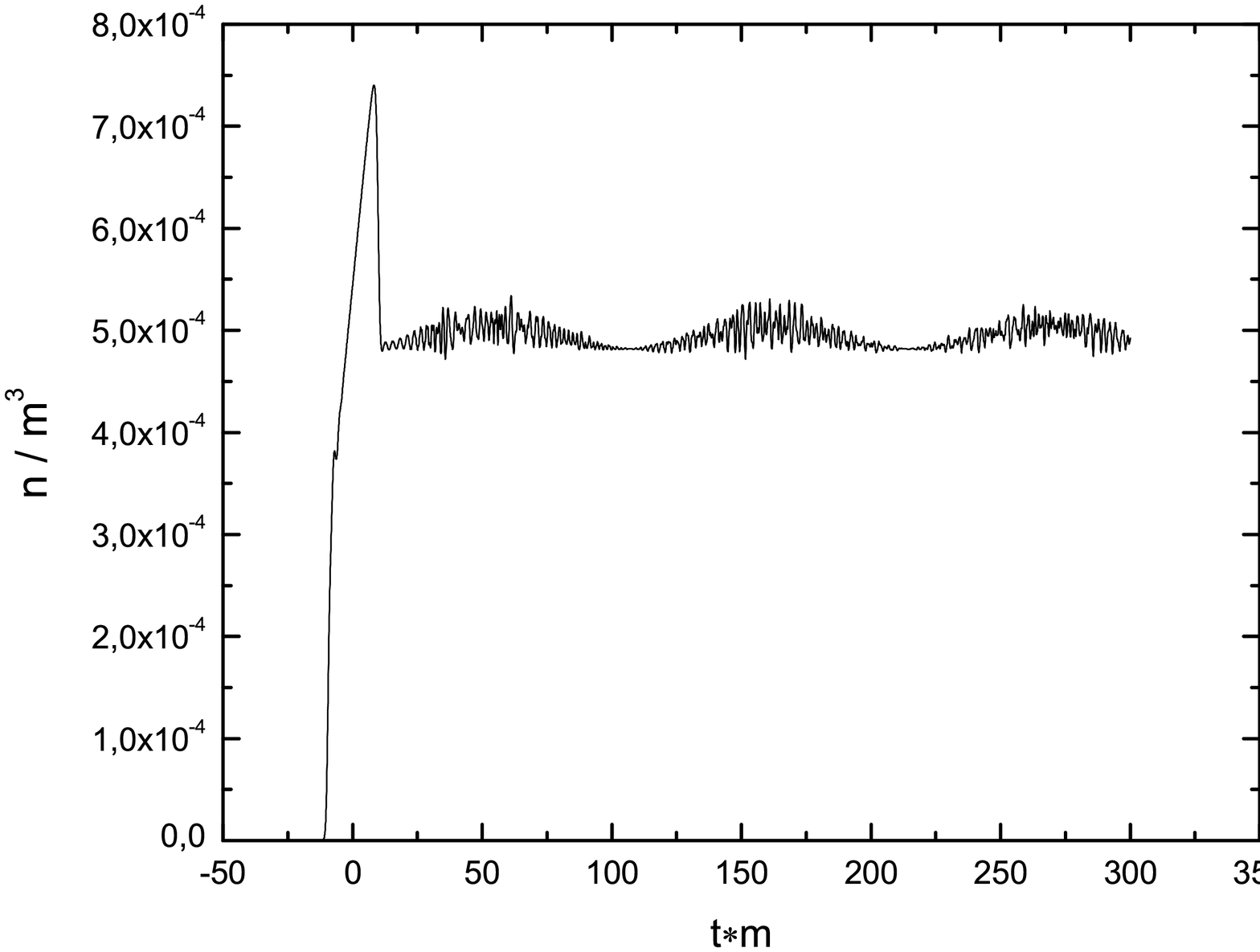}
\vspace*{8pt} \caption{Back reaction in the weak electric and
strong magnetic field ($H=H_{cr}$):
total electric field (left panel), particle
number density (right panel). The stable plasma oscillations are excited
after termination of the external pulse when $E\ge E_{cr}$.}
\end{figure}

\begin{figure}
\centering
\includegraphics[width=55mm,height=45mm]{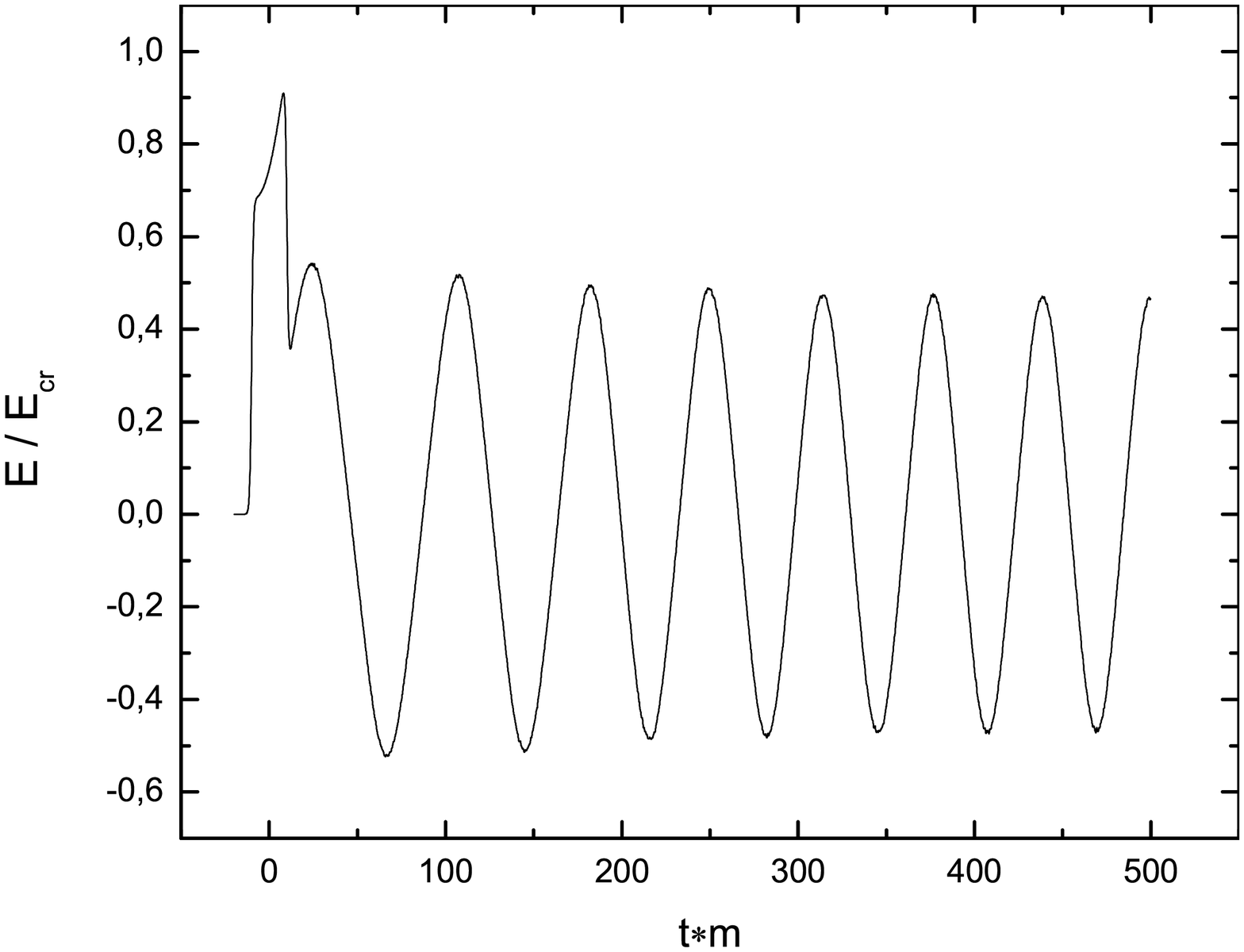}\hspace{5mm}
\includegraphics[width=55mm,height=45mm]{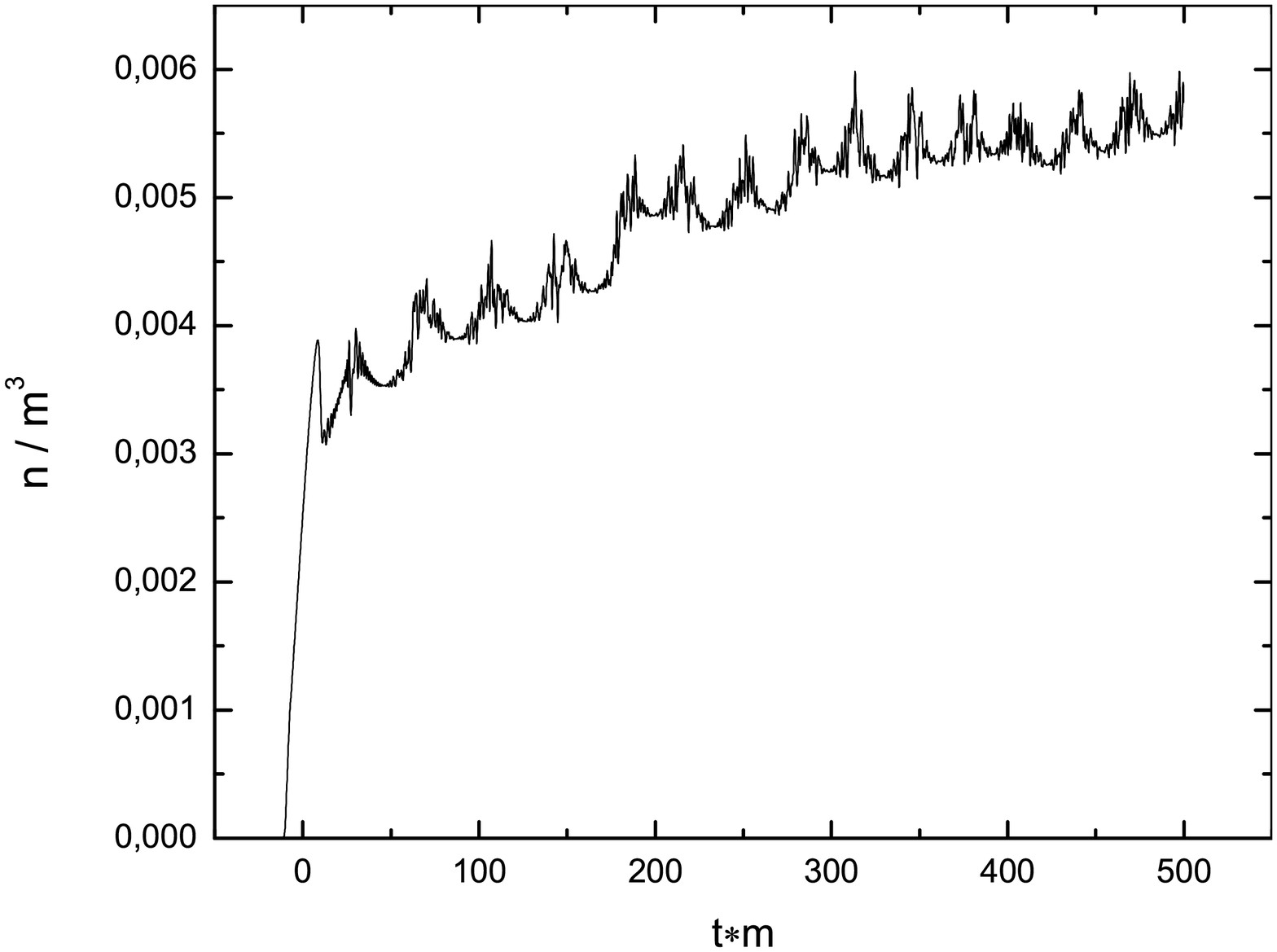}\vspace*{16pt}
\includegraphics[width=55mm,height=45mm]{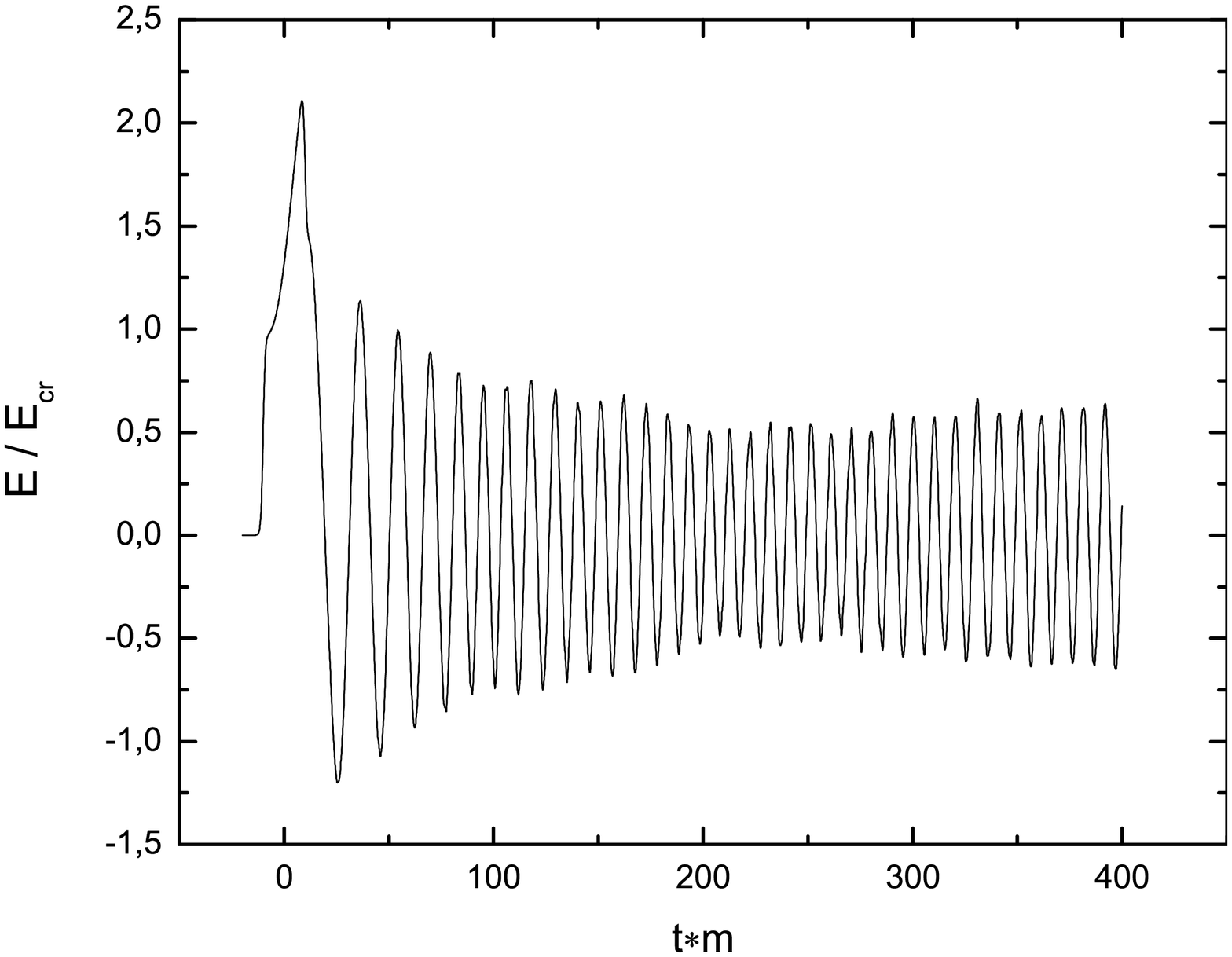}\hspace{6mm}
\includegraphics[width=55mm,height=45mm]{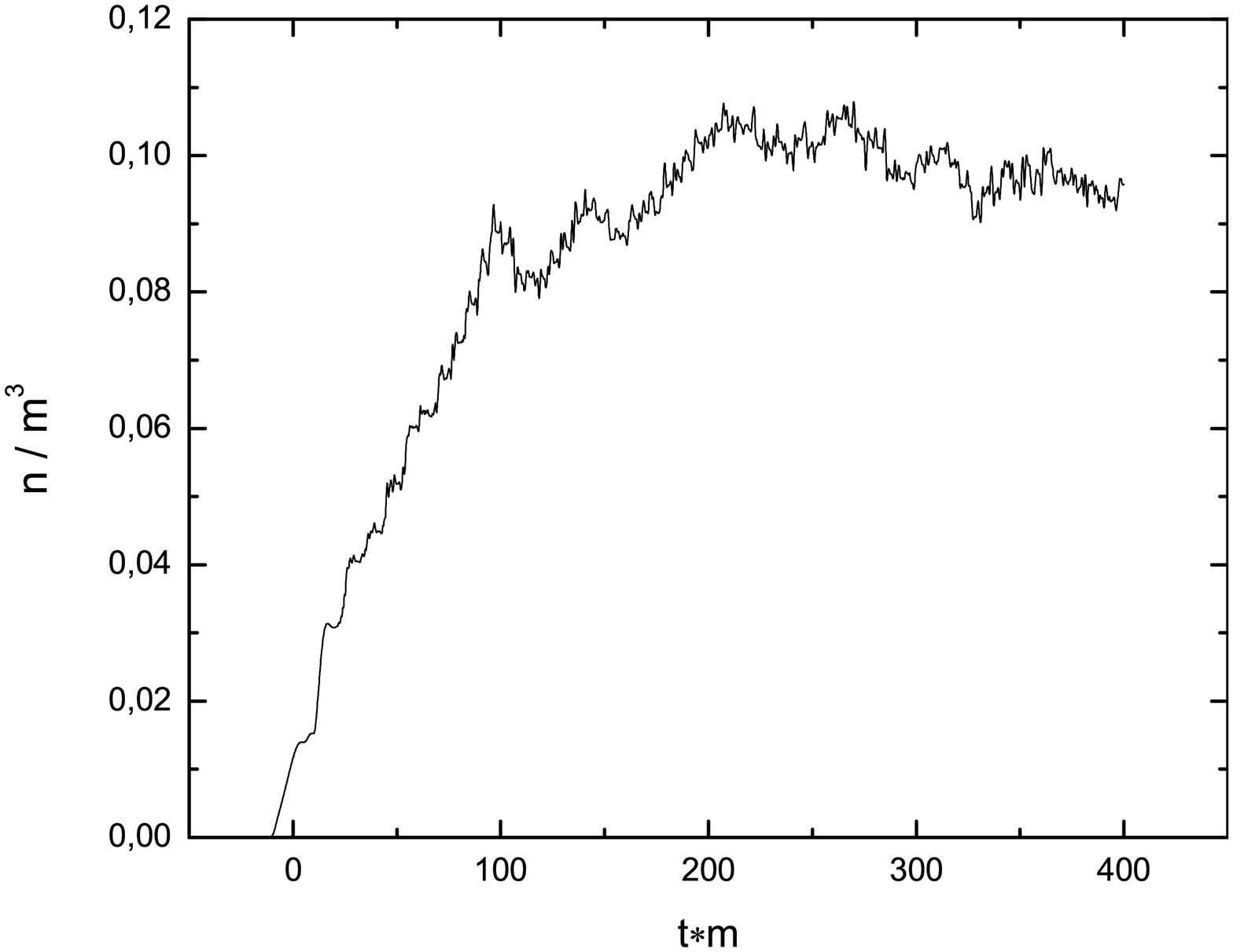}
\caption{Back reaction in the strong electric and magnetic field:
 total electric field (left panel), particle number density (right panel).
 The plasma oscillations becomes large-scale modulated at the increase of
 the field to $E\approx E_{cr}$.}
\end{figure}

During the action of the external pulse \re{rect} with the flat
top, particle density grows linearly. The rate value (without the
magnetic field) is well approximated with the Schwinger formula
for the field strength close to critical value but in the weak
field these quantities  differ strongly. Analogous situation is
for the suppression coefficient of boson creation which is equal
to $\sinh{x}/x$ for $E=const$, where $x=\pi H/E$~\cite{Grib}. For
the top plot on Fig. 2 this value is $\approx 2\cdot 10^{11}$,
whereas the real difference obtained in the kinetic approach is
about 50$\%$.

\section{Summary}

We have shown on an example of scalar QED, the oscillator
(holomorphic) representation is now, apparently, the most
effective non-perturbative method for the kinetic description of
vacuum particle creation. For an illustration, we have obtained
the KE for the case of the arbitrary time-dependent space
homogeneous electric field and applied it for a comparison of the
effectiveness and other features of vacuum tunneling processes in
the linear and circular polarized fields.

Since the corresponding calculations in the spinor QED were not
yet fulfilled, these investigations can be the foundation for the
qualitative understanding of processes of electron-positron plasma
creation in strong laser fields in the planned DESY and SLAC
installations. We have investigated also the influence of the
magnetic field on vacuum particle creation and have studied the
corresponding back-reaction problem, where we have observed great
deviations with respect to prediction of the well known results
for the case of the constant electric and magnetic fields
combination \cite{Grib,niki}.

The results obtained here can be interesting also for the
development of the flux-tube model of superconductive type,
describing the pre-equilibrium evolution of quark-gluon plasma,
generated under conditions of ultrarelativistic heavy ion
collisions \cite{Lam}. The present approach reveals some
perspective in the investigation of the dynamics of vacuum
particle creation in the strong space homogeneous background
fields in more realistic situations when one deals with vacuum
creation of particles with inner degrees of freedom
(electron-positron, quark-gluon plasma etc.).

\section*{Acknowledgements}

This work was supported partly by Deutsche Forschungsgemeinschaft
(DFG) under project number RUS 17/102/00, Russian Federations
State Committee for Higher Education under grant E02-3.3-210 and
Russian Fund of Basic Research (RFBR) under grant 03-02-16877.
The authors are grateful to G.V. Efimov for attention to work
and for an opportunity to familiarize with the book \cite{DEGN}.
One of the authors (V.S.)  is personally grateful  to D.~Blaschke for
the support during the work on this paper.

\appendix

\section{Regularization procedure}

\renewcommand{\theequation}{A.\arabic{equation}}
\setcounter{equation}{0}

The ultraviolet behavior of the unknown functions $f$, $v_1$ and
$v_2$ is determined by the dispersion law $\omega(\ve{p},t)$ and
the vacuum transition amplitude $\Delta(\ve{p},t)$. We can write them
out in the spherical coordinate system and define their asymptotic
in the ultraviolet region for the electric field \re{eq1}
\begin{equation}\label{ea1}
   \Delta(\ve{p},t) = \frac{e\ve{E}(t)\ve{n}q(t)}{\omega^2(q)}
    \mathop{\longrightarrow}_{q\rightarrow\infty}
    \frac{e\ve{E}(t)\ve{n}}{q(t)} \sim \frac{1}{q(t)},
\end{equation}
where $q(t) = \sqrt{\ve{P}^2(t)}$ and $\ve{n} = \ve{P}/q$. Now we
construct the asymptotic series over the inverse powers of
momentum $q$
\begin{equation}\label{ea2}
    f = \sum_{n=1} f^{(n)},\qquad
    v_s = \sum_{n=1} v_s^{(n)}, \qquad s=1,2, \;
\end{equation}
where $f^{(n)}$, $v_s^{(n)} \sim q^{-n}$. We get the following
leading contributions using these decompositions and the Eqs.
\re{ode}
\begin{equation}\label{ea3}
    f^{(4)}(q,t) = \left(\frac{e\ve{E}(t)\ve{n}}{4q^2}\right)^2\! ,
    \quad
    v_1^c=v_1^{(3)}(q,t) = \frac{e\dot{\ve{E}}(t)\ve{n}}{4q^3}, \quad
    v_2^{(2)}(q,t) = \frac{e\ve{E}(t)\ve{n}}{2q^2}.
\end{equation}

The unique purpose of these counter terms is the elimination of
ultraviolet divergence in the integrals defining the densities of
physical quantities (e.g., $v_1^c$ provides the regularization
of the electric current in the r.h.s. of the Eq. \re{eqn34},
$f^{(4)}$ - of the energy density etc.). That is why it is
necessary to introduce some mechanism of switching on of the
counter-terms \re{ea3} only in the asymptotic region $q
\rightarrow \infty$. Switching on the counter terms in the region
$q\rightarrow\infty$ can be achieved, e.g., by the replacement of
$q^2\rightarrow q^2+M^2$ in the Eqs. (A3), where $M^2$ is an
auxiliary  mass parameter (Pauli-Willars procedure), and trending
$M\rightarrow\infty$ after calculations of the integral. Another
possibility is to modify the counter terms, e.g.,
\begin{equation}\label{ea4}
    f_M^{(4)}(q,t) = f^{(4)}(q,t) F_M(q,t)
\end{equation}
with the switching function of the type $F_M(q,t) = 1-
\exp(-q/M)$.

When using the computer calculations and therefore introducing the
cut-off momentum parameter ($\Lambda = q_{max}$ is a grid
boundary), one can always assume counter terms to be negligible
small and omit them.

\end{document}